\documentclass{kluwer}    

\usepackage[dvips]{graphicx}

\begin{document}
\begin{article}
\begin{opening}
\title{Resonant and Secular Families of the Kuiper Belt}
\author{E. I. \surname{Chiang}}
\author{J. R. \surname{Lovering}}
\institute{University of California at Berkeley}
\author{R. L. \surname{Millis}}
\author{M. W. \surname{Buie}}
\author{L. H. \surname{Wasserman}}
\institute{Lowell Observatory}
\author{K. J. \surname{Meech}}
\institute{Institute for Astronomy, Hawaii}

\date{July 11, 2003}
\runningauthor{Chiang et al.}
\runningtitle{Resonant and Secular Families}

\begin{abstract}
We review ongoing efforts to identify occupants of mean-motion
resonances (MMRs) and collisional families in the Edgeworth-Kuiper belt.
Direct integrations of trajectories of Kuiper belt objects (KBOs)
reveal the 1:1 (Trojan), 5:4, 4:3,
3:2 (Plutino), 5:3, 7:4, 9:5, 2:1 (Twotino), and 5:2 MMRs
to be inhabited. Apart from the Trojan, resonant KBOs
typically have large orbital eccentricities
and inclinations. The observed pattern
of resonance occupation is consistent with resonant capture
and adiabatic excitation by a migratory Neptune; however,
the dynamically cold initial conditions prior to resonance
sweeping that are typically assumed by migration
simulations are probably inadequate. Given the dynamically
hot residents of the 5:2 MMR and the substantial inclinations observed
in all exterior MMRs, a fraction of the primordial belt was likely
dynamically pre-heated prior to resonance sweeping.
A pre-heated population may have arisen as Neptune gravitationally
scattered objects into trans-Neptunian space.
The spatial distribution of Twotinos offers a unique diagnostic
of Neptune's migration history.
The Neptunian Trojan population may rival the Jovian Trojan population,
and the former's existence is argued to rule out violent orbital
histories for Neptune. Finally, lowest-order secular theory
is applied to several hundred non-resonant
KBOs with well-measured orbits to update proposals
of collisional families. No convincing family
is detected.
\end{abstract}

\end{opening}

\section{Introduction}
Plutinos are Kuiper belt objects (KBOs) that occupy the
exterior 3:2 mean-motion resonance (MMR) established
by Neptune (see, e.g., Jewitt \& Luu, 2000).
The preponderance of Plutinos having large
orbital eccentricities has been interpreted to imply
that Neptune's orbit expanded outwards by several AUs
over a timescale of $\tau \gtrsim 10^6$ yr (Malhotra, 1995).
The expansion was supposedly driven by angular
momentum exchange with ancient planetesimals interspersed
among the giant planets and having about as much
mass as the ice giants (Fernandez \& Ip, 1984; Hahn \& Malhotra, 1999;
Gomes, 2003). As Neptune spiralled
outwards, its exterior MMRs swept across the primordial
Kuiper belt, captured KBOs, and amplified their orbital
eccentricities.

Simulations for which Neptune's outward
migration is smoothly monotonic and for which
$\tau \gtrsim 10^7$ yr predict the 2:1 MMR to be
populated with roughly as many objects as the 3:2 MMR,
for reasonable assumptions regarding the
distribution of orbital elements prior to resonance
sweeping (Malhotra, Duncan, \& Levison, 2000; Chiang \& Jordan, 2002).
Initial discoveries
of Plutinos but not of ``Twotinos'' (2:1 resonant KBOs)
led to speculation that $\tau < 10^6$ yr;
the strength of the 2:1 MMR is weaker than that of the
3:2, and the former resonance's capture efficiency decreases more rapidly
with increasing migration rate (Ida et al., 2000; Friedland, 2001).
Reports of the absence
of Twotinos proved greatly exaggerated; wide-field surveys for
KBOs and painstaking astrometric recovery observations world-wide
have now secured $\sim$200 KBO orbits with sufficient
accuracy that $\sim$7 Twotinos are confidently identified (Chiang \& Jordan,
2002; Chiang et al., 2003, hereafter C03; and see Table I
of the present paper).
A plethora of other resonances are also observed to be
occupied; what these other resonances imply about the dynamical
history of Neptune and the Kuiper belt is summarized herein.

Following Hirayama (1918), we ask also whether
certain KBOs trace their lineage to
parent bodies that experienced catastrophic, collisional
disruption. The proportion of KBOs that are
shattered fragments records
the collisonal history of the belt and
constrains its mass as a function of time.
Candidate collisional families are identified by similarities
in their observed spectra and in their so-called ``proper''
or ``free'' orbital elements. Here we present a first-cut
calculation of the free elements of KBOs with
accurately measured orbits.

In \S\ref{ri}, we describe our procedure for rigorously
identifying resonant KBOs. In \S\ref{tri}, we highlight
the implications of 3 occupied resonances---the 2:1, 5:2, and 1:1 MMRs---for
the dynamical history of the outer solar system.
In \S\ref{cf}, we present the free orbital elements
of 227 non-resonant KBOs and attempt to identify
a candidate collisional family.

\section{Resonance Identification}
\label{ri}

By definition, a mean-motion resonant KBO is characterized
by one or more resonant arguments that librate, i.e., undergo
bounded oscillations with time. Each resonant argument
takes the form

\begin{equation}
\phi_{p,q,m,n,r,s} = p\lambda - q\lambda_N - m\tilde{\omega} -n\Omega -
r\tilde{\omega}_N - s\Omega_N \; ,
\end{equation}

\noindent where $\lambda$, $\tilde{\omega}$, and $\Omega$
are the mean longitude, longitude of pericenter,
and longitude of ascending node of the KBO, respectively.
Those same quantities subscripted by ``$N$'' are those of Neptune,
and $p$, $q$, $m$, $n$, $r$, and $s$ are integers. By rotational
invariance, $p-q-m-n-r-s=0$.

Identifying resonant objects is a straightforward matter of
integrating forward the trajectories of Neptune and of the KBOs
and examining the behavior of $\phi_{p,q,m,n,r,s}$ for every object.
Our present implementation tests for libration of 107 different
values of $\{p,q,m,n,r\}$; for convenience, and because resonances
associated with the small inclination of Neptune are weak,
we set $s=0$.\footnote{We also test for membership in the secular
Kozai resonance, in which the argument of perihelion, $\omega$,
librates. However, the 3-Myr duration of our integrations is marginally
too short to witness a full Kozai libration cycle for several objects.}
The integrations are carried out with
the SWIFT software package (swift\_rmvs3),
developed by Levison \& Duncan (1994)
and based on the N-body map of Wisdom \& Holman (1991).
We include the influence of the four giant planets, treat each
KBO as a massless test particle, and integrate trajectories
forward for 3 Myr using a timestep of 50 days, starting at
Julian date 2451545.0. Any duration of integration longer than
the mean-motion resonant libration period, $\sim$10$^4$ yr, would be adequate
to test for resonance membership. However, we have found by numerical
experiment that adopting durations less than $\sim$1 Myr yields
membership in a host---often, more than 5---weak resonances for a given
object. Upon integrating for longer durations, many objects escape
most of these high-order resonances. Since we are interested
in long-term, presumably primordial residents of resonances,
we integrate for as long as is computationally practical, i.e., 3 Myr.
In cases of particular interest---e.g., the Neptune Trojan---we
integrate trajectories up to 1 Gyr to test for long-term stability.

Initial positions and velocities of 407 objects
are computed using the formalism of Bernstein \& Khushalani (2000) in the
case of short-arc orbits, and from E.~Bowell's database
in the case of long-arc orbits. These data are maintained
and continuously updated at Lowell Observatory; we report
here results obtained using orbit solutions calculated
on Jan 2 2003. About half of these 407 objects were discovered by the
Deep Ecliptic Survey (Millis et al., 2002; C03; Elliot et al., 2003).
For every object, we integrate forward the best-fit orbit solution
in addition to 2 ``clones'': orbit solutions that lie on the
$3\sigma$ confidence surface and that are characterized
by maximum and minimum semi-major axes. For the cloned
solutions, the other 5 orbital
elements are adjusted according to their correlation with
semi-major axis. Our rationale for singling out
semi-major axis is explained in C03.

\begin{figure}
\centerline{\includegraphics[width=30pc]{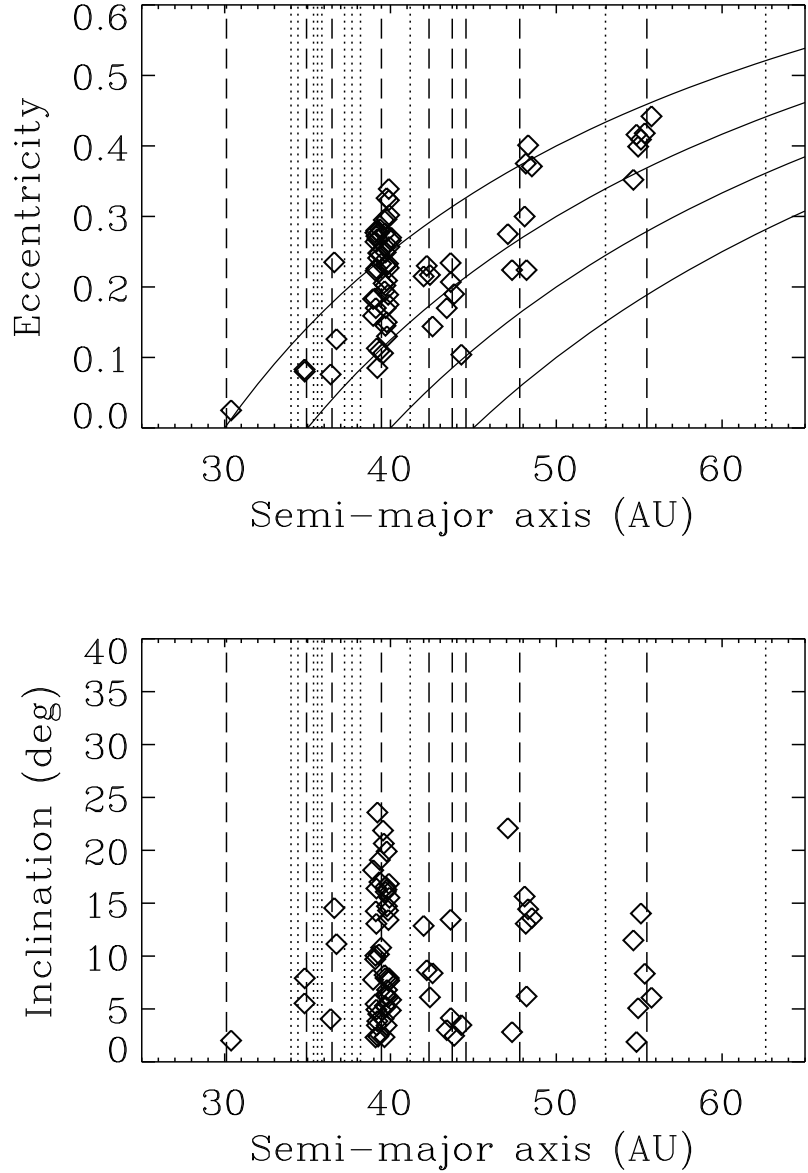}}
\caption[ae0.eps]{Eccentricities, inclinations, and semi-major axes
of 75 securely resonant KBOs. Dotted lines indicate locations of
nominal resonance with Neptune; dashed lines indicate occupied
resonances. In order of increasing semi-major axis, the occupied
resonances include the 1:1 (Trojan), 5:4, 4:3, 3:2, 5:3, 7:4, 9:5, 2:1,
and 5:2 MMRs. Elements are heliocentric, referred
to the J2000 ecliptic plane, and evaluated at epoch 2451545.0 JD.
Uncertainties are too small to plot here.}
\label{aei}
\end{figure}

\begin{figure}
\centerline{\includegraphics[width=25pc]{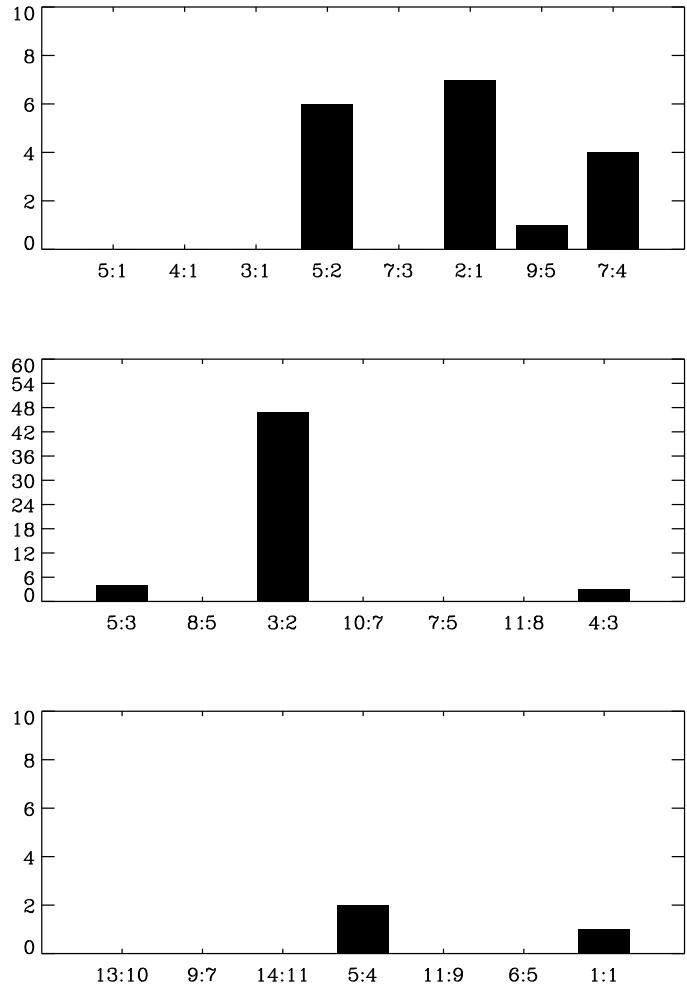}}
\caption[breakerr1.eps]{Number of KBOs occupying a given resonance.}
\label{histo}
\end{figure}

An object is considered ``securely resonant''
if all three sets of initial conditions
yield libration
of the same resonant argument(s)
for the entire duration of the integration. Of 407 KBOs tested,
we find 75 to be securely resonant. Not all of the
407 KBOs have orbital parameters measured
with sufficient accuracy to permit meaningful classification
and more than our 75 identified objects may well be resonant;
a more careful assessment of the resonant vs. non-resonant
population ratio is deferred to Elliot et al.~(2003).
Orbital elements of securely resonant KBOs
are plotted in Figure \ref{aei}. Occupied
resonances include the 1:1 (Trojan),
5:4, 4:3, 3:2, 5:3, 7:4, 9:5, 2:1, and 5:2 MMRs.
They are all of eccentricity-type, though occasionally
objects inhabit both eccentricity-type
and inclination-type resonances.
A histogram of resonance occupation, listing
all resonances that we test for, is displayed
in Figure \ref{histo}. Names of the 75 resonant KBOs
are listed in Table I. The results presented in Figures
1 and 2 and Table 1 differ slightly from those presented in C03,
because here we have made our test for resonance membership
more stringent by adopting a 3$\sigma$ criterion rather
than a 1$\sigma$ criterion; only a handful
of objects ($\sim$10) are found in C03 and not the present work.

\begin{table}
\caption{Resonant 3$\sigma$-confident KBOs}
\begin{tabular}{cc} \hline
Resonance & Name \\ \hline
1:1 & 2001QR$_{322}$ \\
5:4 & 1999CP$_{133}$, 2002GW$_{32}$ \\
4:3 & 1998UU$_{43}$, 2000CQ$_{104}$, (15836) 1995DA$_2$ \\
3:2 & (28978) Ixion$$, 1998UR$_{43}$,
1998WS$_{31}$, 1998WU$_{31}$, 1998WV$_{31}$, \\
    & 1998WW$_{24}$, 1998WZ$_{31}$, 2000CK$_{105}$,
2001KY$_{76}$, 2001KB$_{77}$, 2001KD$_{77}$, \\
    & 2001QF$_{298}$, 2001QG$_{298}$,
2001RU$_{143}$, 2001RX$_{143}$, (15788) 1993SB, \\
    & (15789) 1993SC, (15810) 1994JR$_1$, (15820) 1994TB, (15875)
1996TP$_{66}$, \\
    & (19299) 1996SZ$_4$, (20108) 1995QZ$_9$, (24952) 1997QJ$_4$, (32929)
1995QY$_9$, \\
    & (33340) 1998VG$_{44}$, (38628) 2000EB$_{173}$, (47171) 1999TC$_{36}$,
(47932) 2000GN$_{171}$, \\
    & (55638) 2002VE$_{95}$, 1993RO, 1995HM$_5$, 1996RR$_{20}$, 1996TQ$_{66}$,
1998HH$_{151}$, \\
    & 1998HK$_{151}$, 1998HQ$_{151}$, 1999CE$_{119}$, 1999CM$_{158}$,
1999TR$_{11}$, 2000FV$_{53}$, \\
    & 2000GE$_{147}$, 2001FL$_{194}$, 2001FR$_{185}$, 2001VN$_{71}$,
2001YJ$_{140}$, \\
    & 2001FU$_{172}$, 2002GY$_{32}$ \\
5:3 & (15809) 1994JS, 1999CX$_{131}$, 2001XP$_{254}$, 2001YH$_{140}$ \\
7:4 & 2000OP$_{67}$, 2001KP$_{77}$, 1999KR$_{18}$, 2000OY$_{51}$ \\
9:5 & 2001KL$_{76}$ \\
2:1 & 2000QL$_{251}$, (20161) 1996TR$_{66}$, (26308) 1998SM$_{165}$, \\
    & 1997SZ$_{10}$, 1999RB$_{216}$, 2000JG$_{81}$, 2001FQ$_{185}$ \\
5:2 & (38084) 1999HB$_{12}$, 2001KC$_{77}$, (26375) 1999DE$_9$, \\
    & 2000FE$_8$, 2001XQ$_{254}$, 2002GP$_{32}$ \\
\hline
\end{tabular}
\end{table}

Nearly all of the resonances that are observed to be occupied
are predicted to be occupied by standard, smooth migration
simulations for which $\tau \gtrsim 10^6$ yr. Noteworthy
exceptions include the (innermost) 1:1 and (outermost) 5:2 MMRs.
We turn our attention to these extreme
resonances, in addition to the newly discovered class
of (2:1) Twotinos, to examine their cosmogonic implications.

\section{Three Resonances of Interest}
\label{tri}

\subsection{The 5:2 MMR}
This resonance is difficult to populate under standard
migration scenarios that presume cold initial conditions.\footnote{This
discussion is subject to the caveat that only a fraction
of the observed 5:2 resonant KBOs may be primordial, long-term
residents. C03 undertake a Gyr-long integration
of three observed 5:2 resonant KBOs and conclude
that at least two (38084 and 2001KC$_{77}$) can be primordial
residents with moderate libration amplitudes [$\max(\phi) - 180^{\circ}$]
of less than 100$^{\circ}$. The third KBO (1998WA$_{31}$)
departs the resonance after
3 Myr; it may therefore be a scattered-disk object that has
only temporarily stuck to the 5:2 MMR. Based on this preliminary study,
it seems clear that at least some of the observed objects
are primordial, and to the extent that objects 38084
and 2001KC$_{77}$ have more precisely measured orbits
than 1998WA$_{31}$, the primordial subset may grow as the
astrometry improves.}
C03 report that if $\tau \sim 10^7$ yr,
and if the initial eccentricities and inclinations
of KBOs prior to resonance sweeping are low
($e_{init}, i_{init} \lesssim 0.05$), then the final, post-sweeping
population ratio between the 5:2 and 2:1 resonant objects is
of order 1-to-90. This ratio conflicts
with the observed ratio of $\sim$6-to-7. Moreover,
the simulation predicts final eccentricities
of 5:2 resonant KBOs of $e \sim 0.2$ and final
inclinations of $i \lesssim 1^{\circ}$, values
too small compared with those observed.
Indeed, observed occupants of the 5:2 MMR hold the record
among resonant objects for the highest eccentricities
(up to $e \sim$ 0.45), and they sport high inclinations (up to
$i \sim$ 15$^{\circ}$) as well.

As demonstrated by C03,
observations of 5:2 resonant KBOs may still be reconciled with
the migration hypothesis if one presumes hot initial conditions
prior to resonance sweeping.
C03 find numerically that the sweeping 5:2 MMR more easily captures
KBOs that already possess eccentricities and inclinations
of order 0.2 prior to resonance encounter.
(Of course, by presupposing such highly excited orbits,
the need for any planetary migration at all becomes less pressing!)
The abundance of hot particles in the 5:2 MMR, together
with the large orbital inclinations observed across the entire belt
(both in and out of resonances) and the existence
of high-perihelion objects such as 2000CR$_{105}$ (Millis et al., 2002;
Gladman et al., 2002),
clearly point to at least one other heating mechanism
apart from adiabatic excitation by slowly sweeping MMRs.

What might have caused this pre-heating? We are aware
of two proposals. Thommes, Duncan, \& Levison (2002)
propose that the embryonic cores
of Neptune and Uranus, both of mass $\sim$10 $M_{\oplus}$,
were scattered by Jupiter and Saturn into the ancient belt
and heated KBOs by dynamical friction. Alternatively,
Gomes (2003) points out that under the classic migration scenario,
planetesimals should have undergone close encounters with Neptune
that propelled them onto orbits having larger semi-major axes, eccentricities,
and inclinations, and that
these scattered planetesimals were subsequently swept over
by mean-motion resonances. Our discovery of the
first Neptune Trojan (C03) leads us to favor the mechanism
of Gomes (2003), as we discuss in \S\ref{tro}.

\subsection{The 2:1 MMR}
Twotinos furnish a diagnostic of planetary migration (Chiang \& Jordan, 2002).
The 2:1 resonance allows for asymmetric libration; at large
KBO eccentricities, $\phi_{2,1,1,0,0,0}$ ceases to
librate about $180^{\circ}$, and instead librates about
angles in the vicinity of $\pm 70^{\circ}$ (Beauge, 1994). Whether
a KBO is swept into libration about $\langle \phi \rangle \approx
70^{\circ}$ or into libration about $\langle \phi \rangle \approx -70^{\circ}$
depends on the migration timescale. If $\tau \approx 1$--$3 \times 10^6$ yr,
three times as many KBOs librate about the latter angle
than about the former. The magnitude of the asymmetry monotonically
decreases with increasing $\tau$, and nearly vanishes
if $\tau \gtrsim 10^7$ yr.

\begin{figure}
\tabcapfont
\centerline{%
\begin{tabular}{c@{\hspace{0pc}}c}
\includegraphics[width=2.4in]{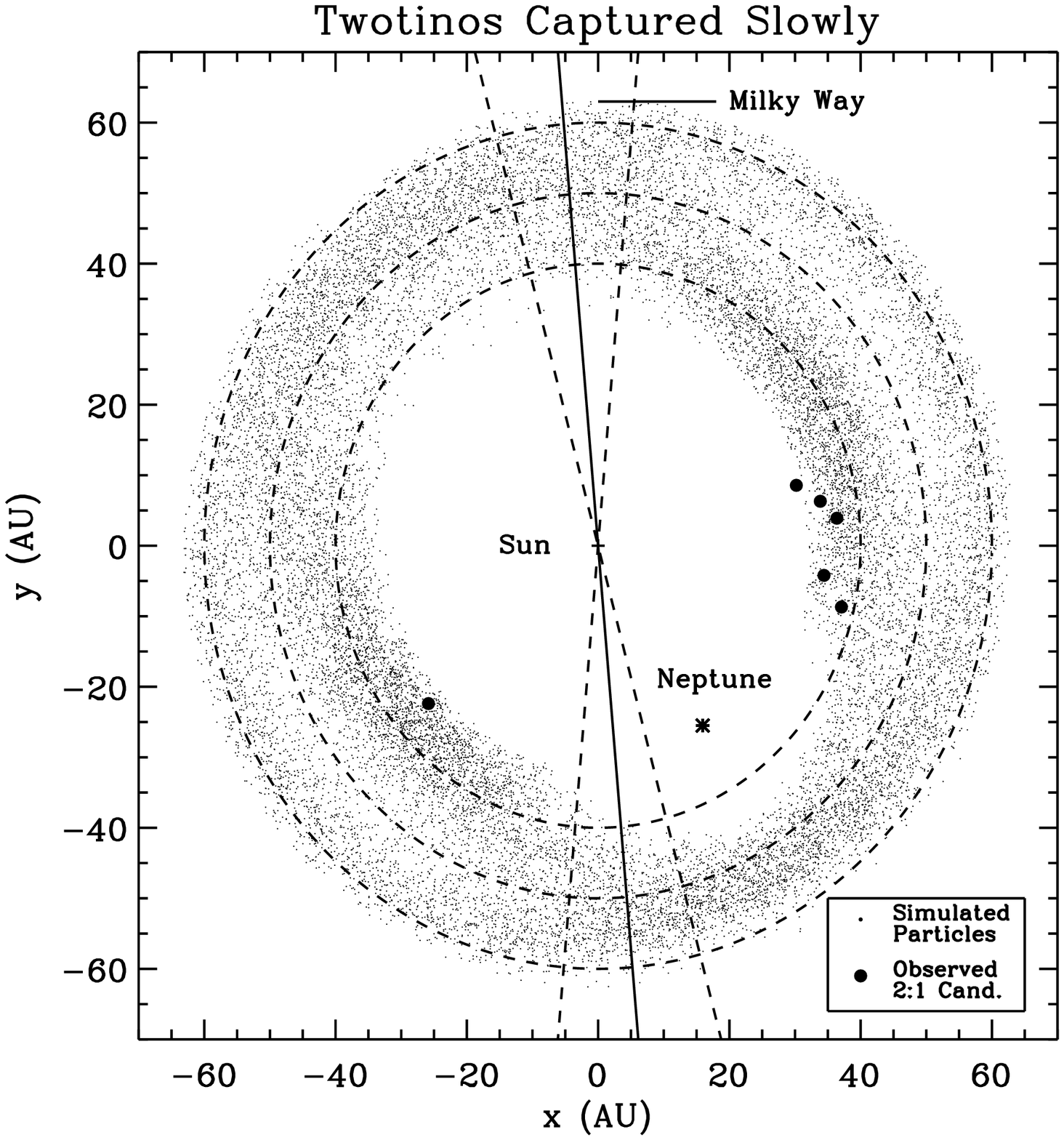} &
\includegraphics[width=2.4in]{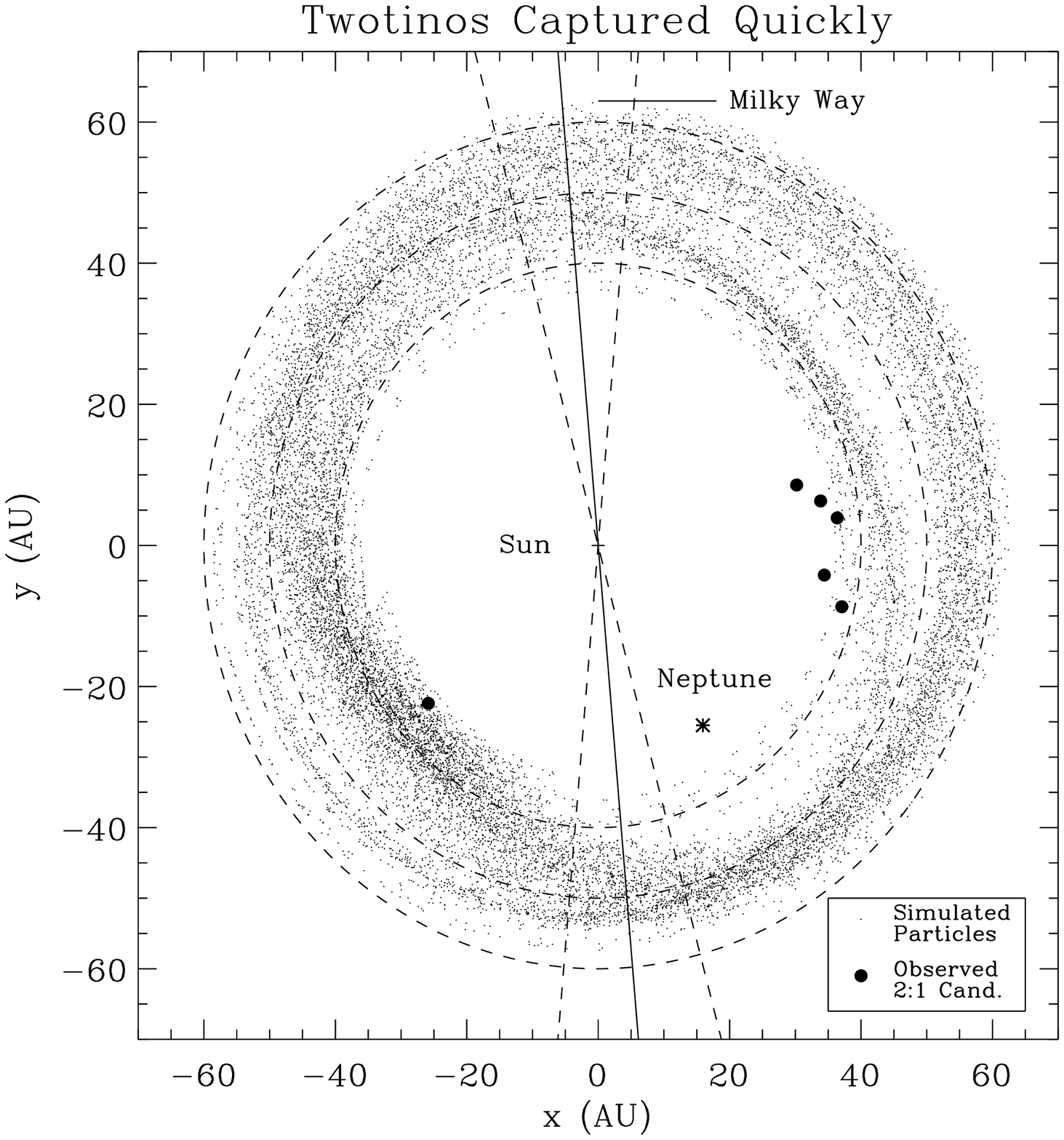} \\
a.~~ $\tau = 10^7$ yr & b.~~ $\tau = 10^6$ yr
\end{tabular}}
\caption{Predicted snapshots, viewed from the ecliptic pole, of
the spatial distribution of 2:1 resonant Kuiper belt objects.
In the left panel, Twotinos were captured into resonance
by Neptune as that planet migrated outwards into the Kuiper Belt over
a timescale of $10^7$ yr. In the right panel, the migration timescale
is $10^6$ yr. Whether the ancient outward migration of Neptune
was slow or fast has dramatic consequences for the longitudinal
distribution of Twotinos. Positions of several recently
discovered Twotinos are marked by solid
circles; they are too few to test these ideas.
Dashed circles correspond to heliocentric radii of 40, 50, and 60 AU,
and radial lines indicate the position of the Galactic plane, $\pm 10^{\circ}$
Galactic latitude.}\label{asym}
\end{figure}

The asymmetry in libration center populations translates directly
into an asymmetry on the sky, as illustrated in Figure \ref{asym}.
If $\tau < 10^7$ yr, more Twotinos should be seen
coming into perihelion at longitudes lagging Neptune's
instantaneous longitude than at longitudes leading it. Observed numbers
of Twotinos are too low to test this prediction.
If such an asymmetry were to be observed in the future
(say, with the PAN-STARRS observatory), it would constitute
strong evidence supporting planetary migration.
If the distribution of Twotinos is found to be symmetric
with respect to the Sun-Neptune line, then such a finding
would be consistent with the planetary migration
hypothesis and would force $\tau > 10^7$ yr.\footnote{The third
and last possibility---that leading Twotinos outnumber lagging
Twotinos---is not predicted at all by the migration model.
If such an observation were to come to pass,
theorists would be forced back to the drawing board.}

\subsection{The 1:1 MMR}
\label{tro}

The first known Neptune Trojan, 2001QR$_{322}$, was discovered by the Deep
Ecliptic Survey (C03). The object can librate about Neptune's forward Lagrange
point (L4) in a tadpole-type trajectory for at least 1 Gyr (C03).
The osculating, heliocentric, and J2000 ecliptic-based eccentricity and
inclination are small, of order 0.03.
The libration center is $\langle \phi_{1,1,0,0,0,0} \rangle \approx
65^{\circ}$, the libration amplitude is $\Delta \phi \equiv \max(\phi) -
\langle \phi \rangle \approx 24^{\circ}$, and the libration period is $T
\approx 10^4$ yr.\footnote{The computed libration center is offset from the
true stable point
of 60$^{\circ}$ because tadpole trajectories are not symmetric
about the Lagrange point.}
For an albedo of 12--4\%, the diameter of 2001QR$_{322}$
is 130--230 km. Based on the area of sky observed to date by the Deep
Ecliptic Survey and various assumed distributions of orbital elements
of Neptune Trojans (see Nesvorny \& Dones, 2002), the total number of Neptune
Trojans resembling 2001QR$_{322}$ ranges between 20 and 60.
Such a population would be comparable to that of Jupiter's Trojans,
for which $\sim$10 exist having diameters between 100 and 200 km
(Davis et al.~2003).

Trojans probably do not owe their existence to planetary
migration; the overwhelming fate of particles that cross
Neptune's orbit is to be scattered onto orbits having high
eccentricities, high inclinations, and semi-major
axes substantially different from Neptune's (C03). One probable
step in the process of accruing Trojans is substantial mass accretion
by the host planet. If the mass of the host planet
grows on a timescale longer than the Trojan libration period,
libration amplitudes of test particles loosely bound
to co-orbital resonances shrink; the planet effectively tightens
its grip as its mass increases.
Horseshoe-type orbits shrink to tadpole-type orbits
(Marzari \& Scholl, 1998), and libration
amplitudes of tadpole-type orbits
further decrease with increasing mass, $M$, of the host planet as

\begin{equation}
\Delta \phi \propto M^{-1/4}
\end{equation}

\noindent (Fleming \& Hamilton, 2000). The weakness of the dependence
of $\Delta \phi$ on $M$ argues that tightening of Trojan orbits
occurred while the host planet accreted the lion's share of
its mass. Thus, we are led to the following picture
for Neptune's formation and orbital evolution. Neptune accreted
the overwhelming bulk of its mass
near a heliocentric distance of $\sim$23 AU on a nearly circular orbit
and, in so doing, captured a Trojan population by
adiabatically securing its hold on whatever co-orbital
planetesimals were present. Subsequent slow migration
of Neptune and the other giant planets whittled down but did not
eliminate Neptune's Trojan population; Gomes (1998) and
Kortenkamp, Malhotra, \& Michtchenko~(2003)
find that standard planetary migration
scenarios reduce the number of Neptune Trojans to a fraction
of order 10\% of their original population. The orbital elements
of surviving Trojans resembles that of long-term stable Trojans as
delineated by Nesvorny \& Dones (2002).

The above picture in which Neptune forms
as the solar system's outermost giant planet core, and in which it
never occupies a substantially eccentric orbit, conflicts
with that of Thommes, Duncan, \& Levison (2002).
In their view, the bulk ($\sim$50\%) of Neptune is assembled
between Jupiter and Saturn; proto-Neptune is subsequently
gravitationally scattered onto a highly eccentric orbit that takes it
into the Kuiper belt. Its trajectory then circularizes
as a consequence of dynamical friction with planetesimals.
We do not understand how Neptune can capture
and retain a retinue of Trojans as it careens back and forth
across the solar system.

\section{Collisional Families}
\label{cf}

We turn now to non-resonant KBOs and ask whether
some objects are collisional fragments based on their
orbital elements. We follow Hirayama (1918) and
compute the free eccentricities and free inclinations
of KBOs. Objects sharing similar values of the free
elements are deemed members of a candidate collisional family.
To extract the free elements,
we employ the secular theory of Brouwer \& van Woerkom (1950)
to subtract the forced elements from the observed
osculating elements.
The procedure is identical to that described
by Chiang (2002); here we update that work by
increasing our sample size to 227 non-resonant KBOs
whose fractional 3$\sigma$ uncertainties in semi-major
axis are less than 6\%, as estimated using the methodology
of Bernstein \& Khushalani (2000).

\begin{figure}
\tabcapfont
\centerline{%
\begin{tabular}{c@{\hspace{0pc}}c}
\includegraphics[width=2.5in]{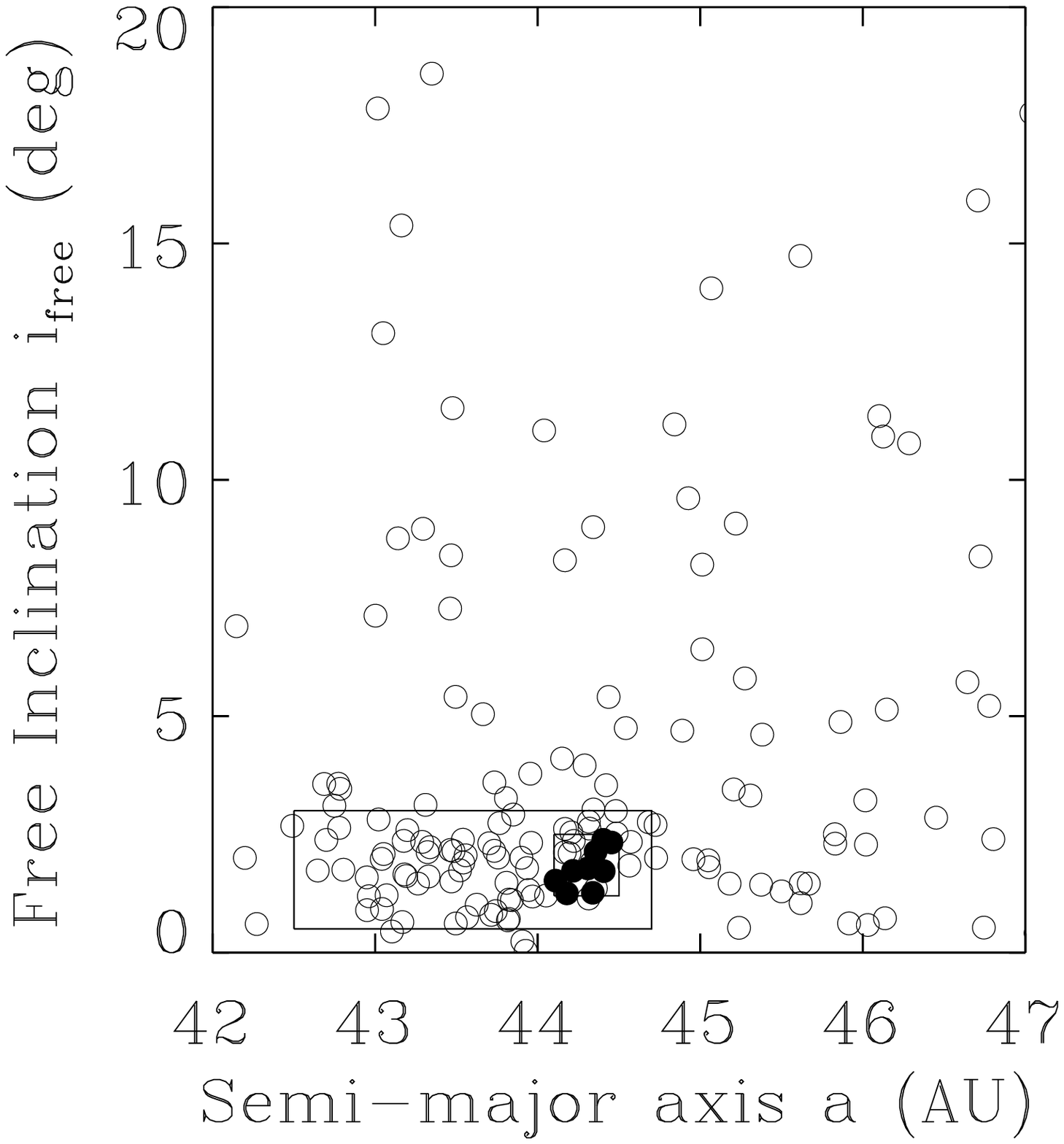} &
\includegraphics[width=2.5in]{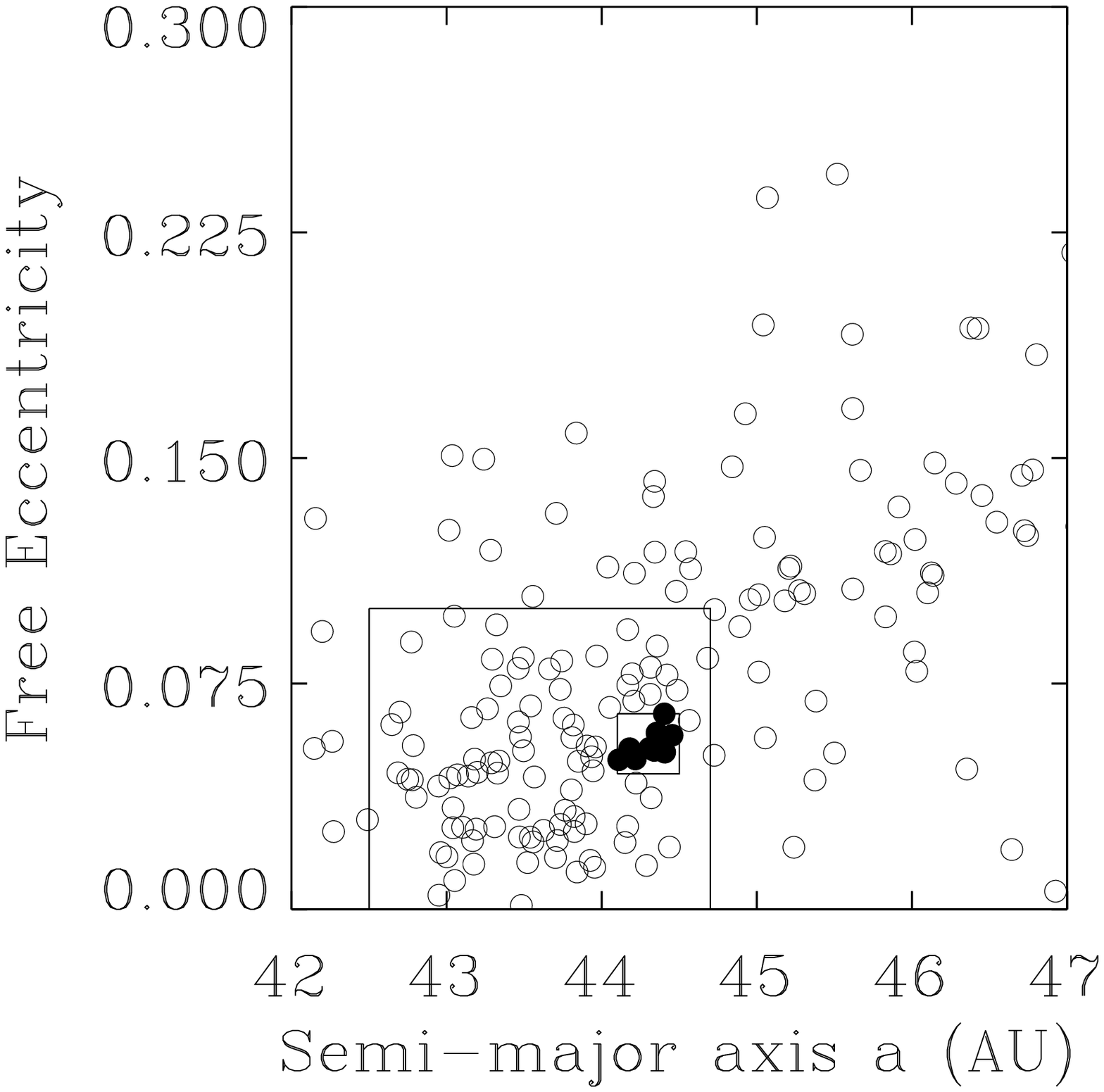} \\
 &
\end{tabular}}
\caption{Free inclinations, free eccentricities, and
osculating semi-major axes of non-resonant KBOs. Solid
circles, enclosed within a small box, mark
9 members of a candidate collisional family. A larger box
is (somewhat arbitrarily) drawn around the volume in which
points seem to be distributed uniformly and encloses 71 points.
If points are distributed randomly within this larger volume
according to a uniform probability distribution,
the probability of 9 points out of 71 lying within the smaller
volume is remarkably small, $\sim$$10^{-6}$. Unfortunately,
the velocity dispersion of the cluster is too small
compared to the escape velocity of the hypothesized parent
body; the cluster of points probably does not correspond
to a real collisional family.}\label{free}
\end{figure}

Figure \ref{free} displays the free eccentricities, free inclinations,
and osculating semi-major axes (which are constants of the motion
in secular theory) of our sample. Nine KBOs are highlighted
that appear, by eye, to be strongly clustered
in ($e_{\rm{free}}$, $i_{\rm{free}}$, $a$) space.
A box having dimensions that can just enclose these nine points,
if placed anywhere else in ($e_{\rm{free}}$, $i_{\rm{free}}$, $a$) space,
encloses fewer than nine points. We list the properties
of the nine KBOs in Table II.
Are these nine KBOs fragments of a once disrupted parent body?
The short answer is, probably not. In what follows,
we describe our efforts at determining the
significance of this clump of points. We offer
arguments for and against the reality of this candidate
family, partly to illustrate the difficulties involved
in identifying real families.

Our candidate family is similar to the one
proposed by Chiang (2002); indeed, three members
are shared between them (1998HM$_{151}$,
1999RC$_{215}$, and 2000PY$_{29}$).
We regard our candidate family to supersede that
proposed by Chiang (2002), since our dataset
is larger and more current. Note that
unlike the family originally proposed
by Chiang (2002), which clusters only
in $a$ and $i_{\rm{free}}$, our candidate family clusters
in all three dimensions. Moreover, the greater size
of our sample now makes clear that
not all of the KBOs in the range of semi-major
axes spanned by our family are probably members of the same family;
additional, less clustered objects exist
at large inclinations and a variety of eccentricities.
This feature lends further support to the
reality of our proposed family. If the objects in Table II
do constitute fragments of the same parent body,
the minimum diameter of the parent body would be
700 km, based on the measured $H_V$'s
and an assumed albedo of 5\%.

We perform three tests to assess the statistical
and physical significance of our candidate family.
The candidate passes the first
test, but fails the other two.

The first {\it ad hoc} and crude estimate
of the statistical significance of this cluster proceeds
as follows. As shown in Figure \ref{free},
we draw a large box that encloses a volume in
which points appear to be distributed
uniformly. There are 71 points within this
large box. Within this volume we draw another, smaller
box that encloses the 9 KBOs. We then ask, if
we randomly distribute 71 points in the larger box
according to a uniform probability distribution,
what is the probability that 9 points out of 71 land within
the smaller box? The answer is $\sim$$10^{-6}$,
a number that we regard to be sufficiently small
to warrant further investigation.

\begin{table}
\caption{Nine Clustered KBOs}\label{tabl}
\begin{tabular}{lcccc} \hline
Name & $a$ (AU) & $e_{\rm{free}}$ & $i_{\rm{free}}$ (deg) & $H_V$ (mag) \\
\hline
(52747) 1998HM$_{151}$ &  44.18 & 0.053 & 1.25 & 7.9 \\
1999OA$_{4}$ & 44.45 & 0.058 & 2.33 & 7.9 \\
1999RC$_{215}$ & 44.40 & 0.065 & 2.38 & 6.9 \\
2000PM$_{30}$ & 44.11 & 0.050 & 1.52 & 7.9 \\
2000PW$_{29}$ & 44.22 & 0.050 & 1.73 & 8.2 \\
2000PY$_{29}$ & 44.34 & 0.053 & 1.26 & 7.1 \\
2000YA$_{2}$  & 44.41 & 0.052 & 1.72 & 6.9 \\
2001QS$_{322}$ & 44.31 & 0.054 & 1.78 & 5.7 \\
2001QZ$_{297}$ & 44.36 & 0.059 & 2.13 & 6.3 \\
\hline
\end{tabular}
\end{table}

A second test, suggested to us by Renu Malhotra,
asks whether the dispersion of ``free velocities''
exhibited by candidate family members matches the expected
dispersion from a catastrophic collision.
A minimum estimate for the latter is the escape velocity of the
parent body; for our putative parent body of minimum
diameter 700 km, the escape velocity is at least
$\sim$0.4 km/s. Immediately after
the hypothesized collision, fragments must have
been moving relative to each other with
velocities near or above this value to avoid
gravitational re-accumulation.  We estimate the actual
velocity dispersion by calculating the standard deviation of
$\sqrt{e_{\rm free}^2 + i_{\rm free}^2} v_K$, where
$v_K \approx 5$ km/s is the Keplerian orbital velocity
of family members. The answer is 0.03 km/s $\ll 0.4$ km/s.
This finding casts doubt on the reality of our proposed family.
However, adding more objects at greater values
of $e_{\rm free}$ and $i_{\rm free}$ to our candidate family
would help to reconcile the velocity dispersions.

A third test, suggested to us by Brad Hansen,
employs Ward's minimum-variance method for
quantitatively identifying
clusters in data sets (Murtagh \& Heck, 1987).
This method agglomerates objects in order of increasing
separation in ($e_{\rm{free}}$, $i_{\rm{free}}$, $a$) space.
A convincing segregation would demand the
nine candidate family members to be agglomerated consecutively
together and the distance between
this agglomeration and others to be large. Unfortunately, not only were
the nine members not agglomerated consecutively together, but no single
agglomeration of objects emerged that was clearly distinguishable
from the remaining data set.

We conclude that no rigorously defensible collisional
family can be identified among the 227 non-resonant
KBOs tested. The tests served to highlight
the subjective nature of identifying families.
Despite the difficulties involved, we emphasize
that a definitive measurement of the proportion
of KBOs that are shattered fragments would
offer direct insight into the belt's mass
and velocity dispersion as a function of time.
If recent proposals regarding the formation of KBO binaries
are correct, so that nearly all KBOs form as nearly equal-mass
binaries (Goldreich, Lithwick, \& Sari 2003),
then KBOs that are found today not to be binary
would comprise the shattered population.

\acknowledgements
We thank the organizing committee of this conference and
the city of Antofagasta for a useful and memorable meeting.
We are indebted to Brad Hansen and Renu Malhotra for sharing
their suggestions on how to test the significance of
candidate families. Support for E.I.C. and J.R.L.
was provided by NSF grant AST-0205892
and the UC Berkeley URAP foundation.

\end{article}
\end{document}